# Tailoring surface topographies on solids with Mid-IR femtosecond laser pulses

S. Maragkaki[1*], G. D. Tsibidis[1,2*], L. Haizer[3], Z. Pápa[3], R. Flender[3], B Kiss[3], Z. Márton[3] and E. Stratakis[1,4*]

[1] *Institute of Electronic Structure and Laser (IESL), Foundation for Research and Technology (FORTH), N. Plastira 100,*
*Vassilika Vouton, 70013, Heraklion, Crete, Greece*

[2] *Department of Material Science and Technology, University of Crete, 71003 Heraklion, Greece*

[3] *ELI-ALPS, ELI-HU Non-Profit Ltd., Wolfgang Sandner utca 3., Szeged, H-6728, Hungary*

[4] *Department of Physics, University of Crete, 71003 Heraklion, Greece*

[*]*Corresponding authors: marag@iesl.forth.gr (S. Maragkaki), tsibidis@iesl.forth.gr (G.D. Tsibidis), stratak@iesl.forth.gr (E. Stratakis)*

## Abstract

Irradiation of solids with ultrashort pulses using laser sources in the mid-infrared (mid-IR) spectral region is a yet predominantly unexplored field that opens broad possibilities for efficient and precise surface texturing for a wide range of applications. In the present work, we investigate both experimentally and theoretically the impact of laser sources on the generation of surface modification related effects and on the subsequent surface patterning of metallic and semiconducting materials. Through a parametric study we correlate the mid-IR pulsed laser parameters with the onset of material damage and the formation of a variety of periodic surface structures at a laser wavelength of $\lambda_L$=3200 nm and a pulse duration of $\tau_p$=45 fs. Results for nickel and silicon indicate that the produced topographies comprise both high and low spatial frequency induced periodic structures, similar to those observed at lower wavelengths, while groove formation is absent. The investigation of the damage thresholds suggests the incorporation of nonlinear effects generated from three-photon-assisted excitation (for silicon) and the consideration of the role of the non-thermal excited electron population (for nickel) at very short pulse durations. The results demonstrate the potential of surface structuring with mid-IR pulses, which can constitute a systematic novel engineering approach with strong fields at long-wavelength spectral regions that can be used for advanced industrial laser applications.

# A.     Introduction

In the past decades, micro- and nano-structuring of materials with ultrashort laser pulses has proven its significance and contribution to major advances in science, technology and industry [1-6]. The fabrication of so-called Laser Induced Periodic Surface Structures (LIPSS) represents an innovative processing approach performed through the femtosecond laser excitation of solid surfaces in conditions close to the ablation threshold [1, 7]. In principle, these self-assembled textures are usually classified according to structure size. Hence, the period of an induced structure $\Lambda$ may range from (i) deep/shallow subwavelength ($\Lambda << \lambda_L/2$, $\lambda_L$ is the laser wavelength), coined as High Spatial Frequency LIPSS (HSFL) [7, 8] through (ii) subwavelength ($\lambda_L/2 < \Lambda < \sim \lambda_L$), termed as Low Spatial Frequency LIPSS (LSFL) [9, 10], to (iii) suprawavelength ($\Lambda > \lambda_L$), called grooves [11-14] or (iv) spikes [13, 15]. In general, the size and complexity of the induced surface patterns can be tailored through the appropriate control of various laser parameters such as fluence, energy dose, polarization states, angle of incidence and pulse separation (i.e. if a train of pulses is used) [1, 16, 17]. The aim is to fabricate topographies of enhanced complexity and features that allow the patterned material to demonstrate remarkable biomimetic functionalities [1].

Another important parameter that also considerably influences surface topographies is the photon energy of the laser source, which is directly related to the laser wavelength [18]. The impact of the photon energy is closely associated to the attained excitation levels. Various experimental and theoretical studies have reported the wavelength dependence of the damage threshold of solids (especially, semiconductors or dielectrics) [19-29]. This irreversible surface modification in materials is related with the generation of nonlinear processes potentially affecting the minimum fluence that leads to material damage [12, 20, 21, 30-33]. In principle, laser sources in a spectral region between the visible and near-infrared frequencies (<1.5 μm) have been employed to explore both ultrafast dynamics and laser-based patterning [1, 32, 33].

However, exciting opportunities with the employment of laser pulses in the mid-IR spectral region have arisen in various fields recently. The motivation and the interest have been generated by the impressive physical phenomena that characterize the response of the material at mid-IR. More specifically, photon energies lower than the band gap of silicon (~1.1 eV) can be used to trigger nonlinear absorption processes in the laser focal volume without producing absorption in the surrounding material and therefore, selectively, inscribe optical waveguides in crystalline silicon [34]. Some other phenomena are related to photon acceleration in metasurfaces, megafilamentation in atmosphere [35], transparency of materials at large $\lambda_L$ [36], significance of Kerr effects [37, 38] or the generation of weakly bound surface plasmons [39]. The above physical phenomena appear to provide numerous possibilities for both fundamental and applied research [36].

Laser-material processing at mid-IR is still a rather unexplored field due to the mainly limited availability of laser sources operating at such long wavelengths. Although, there have been numerous reports on the investigation of the irradiation of solids with mid-IR pulses (see [37, 40, 41] and references therein), only few of them present fabrication techniques of HSFL or LSFL topographies via laser patterning at large $\lambda_L$ [35, 36, 42-44] to . On the other hand, various theoretical studies on the impact of the nonlinear effects produced at those photon energies revealed several remarkable physical phenomena, especially for wide band gap materials [45] or semiconductors [46]. These include surface plasmon (SP) excitation at lower fluences, SP of longer lifetime and larger damping length and weaker confinement, shift of tunnelling assisted material ionization mechanisms at lower energies and influence of impact ionization in damage threshold at longer wavelengths. This abundance of unravelling phenomena in laser-matter interaction can potentially be exploited to benefit patterning approaches for the development of advanced tools in nonlinear optics and photonics for a large range of applications [47].

Nevertheless, there are still several open questions that require further exploration in order to improve the current understanding of the fundamental physical mechanisms and to develop a consistent methodology for tailoring surface topographies with the use of mid-IR sources in a controlled way. The experimental investigation of pattern formation has hitherto been limited, mainly to semiconductors, where HSFL and LSFL were observed [43, 44]. However, we should know whether the nonlinearities introduced by mid-IR pulses can influence the fabrication conditions of these structures, whether grooves and spikes are produced (as in IR), and whether the transition from one type of structure to another can systematically be predicted. On the other hand, to the best of our knowledge, no previous work on LIPSS fabrication on metals with mid-IR sources has been reported.

To address the aforementioned questions, a comparative study and investigation of LIPSS formation in two types of materials (silicon and nickel) was performed to reveal the impact of the underlying nonlinear effects that account for the surface patterning at mid-IR.

## B. Experimental protocol

The experiments were conducted at the Extreme Light Infrastructure Attosecond Light Pulse Source (ELI ALPS). The Mid-IR laser system provided linearly polarized pulses with a wavelength of 3200 nm with 45 fs duration at a repetition rate of 100 kHz [48]. The continuous power attenuation of the mid-IR beam was realized by a pair of holographic wire grid polarizers. By using a gold coated off-axis parabolic mirror with a reflected focal length of 50 mm, the 45 fs mid-infrared pulses were focused to ($1/e^2$) spot diameter of approximately 50 μm with an incident laser beam (Z direction) perpendicular to the surface of the sample (XY plane). The samples were mounted onto a two-axis (XZ plane) translation stage mounted onto a computer-controlled fast translation stage (Y direction). The number of laser shots ($N$=22 to 1000) on a given area was controlled by the speed of the motorized linear stage. Using this approach, a series of lines were patterned upon systematic variation of the input laser power (fluence) and the number of pulses. The morphology of the laser-processed surfaces is characterized via scanning electron microscopy (SEM). A two-dimensional fast Fourier transformation (2D-FFT) of top-view SEM images is performed to determine the periodicities of the surface patterns, using the Gwyddion software.

## C. Theoretical model

To explore the impact of mid-IR laser pulses on solids upon irradiation, it is important to describe the dynamics of the excited carriers, evaluate the induced thermal effects and investigate the underlying surface modification processes. A multiscale approach needs to be employed to couple all physical mechanisms at different time scales in order to interpret the laser pulse fingerprint on the solid's surface.

### I. Ultrafast Dynamics

It is known that excitation processes are different in metals and semiconductors, and therefore, a distinctly different analysis of the impact of pulses at mid-IR for the two materials has to be carried out. In the next sections, we emphasize the presentation of the model simulating the ultrafast dynamics in metals, as a systematic investigation of excitation and the thermal response of semiconductors at mid-IR was presented in detail in a previous report [46]. To compare with experimental results, the simulations were performed at $\lambda_L$=3200 nm laser wavelength and $\tau_p$=45 fs pulse duration.

#### a. Nickel

To describe the ultrafast dynamics of the produced excited carriers and the thermal response of a metallic surface following irradiation with femtosecond laser pulses, a revised version of the traditional Two Temperature Model (rTTM) [49-51] is used in this work. More specifically, due to the extremely short length of the pulse (~45 fs), the traditional Two Temperature Model (TTM) [52] is known to overestimate the amount of the internal energy of thermal electrons due to the formation of an out-of-equilibrium nonthermal electron population produced during the irradiation (see [49] and references therein for alternative approaches). The rTTM constitutes an approach that describes adequately ultrafast dynamics in noble metals [53, 54]; compared to TTM, rTTM (Eq.1) comprises two 'source terms' to describe (i) the electron excitation due to the laser source $\frac{\partial U_{ee}}{\partial t}$, and (ii) the interaction of a nonthermalized electron system both with the thermalized electron system and the lattice $\frac{\partial U_{eL}}{\partial t}$

$$\begin{aligned} C_e \frac{\partial T_e}{\partial t} &= \vec{\nabla} \cdot (k_e \vec{\nabla} T_e) - G_{eL}(T_e - T_L) + \frac{\partial U_{ee}}{\partial t} \\ C_L \frac{\partial T_L}{\partial t} &= G_{eL}(T_e - T_L) + \frac{\partial U_{eL}}{\partial t} \end{aligned} \quad , \tag{1}$$

where $C_e$ and $C_L$ stand for the heat capacity of the electron and lattice systems, respectively, $G_{eL}$ is the electron-phonon coupling and $k_e$ corresponds to the electron's thermal conductivity (expressions for the energy densities $\frac{\partial U_{ee}}{\partial t}$ and $\frac{\partial U_{eL}}{\partial t}$ are presented in [49, 50] and the references therein).

**b. Silicon**

As the photon energy (~0.5 eV) corresponding to $\lambda_L$=3200 nm is smaller than the value of the indirect band gap of silicon ($E_g$~1.12 eV at room temperature), linear single photon or nonlinear two-photon interband absorption processes are not sufficient to move carriers from the valence to the conduction band. Hence, three-photon absorption is required to generate interband transitions. By contrast, the only linear process involved is the free carrier absorption, which causes intraband transitions that, although increase the carrier energy, do not alter the carrier number density [55]. The energy transfer between the carriers (electron-hole) and the lattice is described by the following rate equations

$$\begin{aligned} C_c \frac{\partial T_c}{\partial t} &= \vec{\nabla} \cdot (k_c \vec{\nabla} T_c) - G_{cL}(T_c - T_L) + S(\vec{r}, t) \\ C_L \frac{\partial T_L}{\partial t} &= \vec{\nabla} \cdot (k_L \vec{\nabla} T_L) + G_{cL}(T_c - T_L) \end{aligned} \quad , \tag{2}$$

where $C_c$ and $C_L$ stand for the heat capacities of the carriers and the lattice, respectively, $G_{cL}$ is the carrier-phonon coupling, $k_L$, ($k_c$) correspond to the lattice's (carrier's) thermal conductivity and $S$ stands for a generalized 'source' term [46]. Furthermore, another rate equation is introduced to describe the evolution of the carrier density ($N_c$) of the excited carriers

$$\frac{\partial N_c}{\partial t} = \frac{\gamma_{TPA} I^3}{3\hbar \omega_L} - \gamma N_c^3 + \theta N_c - \vec{\nabla} \cdot \vec{J} \quad , \tag{3}$$

where $\vec{J}$ is the current carrier density, $\gamma$ is the coefficient for Auger recombination ($\gamma$~$10^{-32}$ cm$^6$/s [63]), $\gamma_{TPA}$ is the three-photon absorption coefficient [41, 46] and $\theta$ is the impact ionization rate coefficient. In this work, an intensity dependent expression is used for the impact ionization parameter $\theta$ [33, 64] unlike an approximating formula used for other semiconductors and at lower laser wavelengths, i.e. ~$e^{-E_g/k_B T_c}$ s$^{-1}$ for silicon at 800 nm [22, 57], where $k_B$ is the Boltzmann constant

$$\theta = \frac{e^2 \tau_c I}{c \varepsilon_0 n m_r E_g (\omega_L^2 \tau_c^2 + 1)(2 - m_r/m_e)} \quad . \tag{4}$$

In Eq.4, $c$ is the speed of light, $e$ is the electron charge, $\varepsilon_0$ stands for the vacuum permittivity, $m_r$=0.18 $m_e$ is the reduced electron mass ($m_e$ is the electron mass) and $n$ is the refractive index of the material, while $I$ is the peak intensity of the laser beam, respectively, and $\tau_c$ is the electron collision time ($\tau_c$~1.1 $fs$). A detailed description of the parameters and values for modelling the response of silicon at mid-IR are reported in [46], while an experimental investigation for deep silicon amorphization at mid-IR validates the theoretical predictions of the model at long wavelengths [56].

**II. Fluid transport**

One of the predominant factors that lead to surface modification is a phase transition the material undergoes upon relaxation. More specifically, if the laser conditions are sufficient to raise the lattice temperature to exceed the melting point of the solid, hydrothermal effects are expected to generate a mass displacement of the produced fluid which, in turn, leads to a non-flat surface relief. To quantify the phase transition and resolidification process, a detailed investigation of the induced fluid movement and pattern formation process is required. Assuming the molten material behaves as an incompressible fluid, the dynamics of the molten volume is described by the following Navier-Stokes equation (NSE)

$$\rho_0 \left( \frac{\partial \vec{u}}{\partial t} + \vec{u} \cdot \vec{\nabla} \vec{u} \right) = \vec{\nabla} \cdot \left( -P + \mu (\vec{\nabla} \vec{u}) + \mu (\vec{\nabla} \vec{u})^T \right) \tag{5}$$

where $\rho_0$ and $\mu$ stand for the density and viscosity of molten material, while $P$ and $\vec{u}$ are the pressure and velocity of the fluid [10, 16, 45]. In Eq. 5, superscript $T$ denotes the transpose of the vector $\vec{\nabla} \vec{u}$ [10].

Another process associated with surface patterning is related to mass removal (ablation). In principle, that process is usually modelled by removing lattice points whose temperature exceeds the thermodynamic critical temperature or the boiling point [10, 45, 57, 58] while the rest of the material is treated depending on whether it is in the molten or solid phase.

## D. Results and Discussion

**I. Damage Threshold**

To evaluate the (experimentally) determined damage threshold dependence as a function of the number of pulses $N$, $\Phi_{thr}^{(exp)}(N)$, a systematic analysis is performed through measuring the width of the ablated lines on both nickel (Fig.1a) and silicon (Fig.1b) at various energy doses. The width of the ablated line shown in Figure 1a and 1b corresponds to the mean value of five different measurements performed using an analysis with *ImageJ* software on the SEM images along the ablated line. The error bars (which are not visible because they are very small) are the results of the deviation of each measurement from the mean value. The number of pulses per spot was controlled by choosing the repetition rate and the moving speed of the stage. The total number of the laser pulses $N$, incident to the surface was varied from $N = 22$ to $N = 1000$ pulses per spot area. Each laser-processed surface was fabricated at a constant peak fluence $\Phi_0$ ranging from 0.05 to 10 J cm$^{-2}$. The ablation threshold fluences were determined by an analysis of the results presented in a semi-logarithmic plot that illustrates the scanning line width $D$ versus the peak fluence for a fixed number of pulses per spot $N$, according to the relation: $D^2 = 2\omega_0^2 \ln\left(\frac{\Phi_0}{\Phi_{thr}}\right)$. The extrapolation to $D^2 = 0$ delivers $\Phi_{thr}$ for a constant number of pulses (Figure 1a for nickel and in Figure 1b for silicon). Assuming the presence of incubation effects and fitting the experimental data through the expression $\Phi_{thr}^{(exp)}(N) = \Phi_{thr}^{(exp)}(N = 1) N^{\xi - 1}$, where $\Phi_{thr}^{(exp)}(N = 1)$ stands for a single shot damage threshold and $\xi$ is the incubation parameter [59-61], an estimation of the incubation parameter $\xi$ and experimental value for $\Phi_{thr}^{(exp)}(N = 1)$ are deduced for nickel ($\xi$=0.90±0.1 and $\Phi_{thr}^{(exp)}(N = 1) = 0.2 \pm 0.1$ J/cm$^2$) (Fig.1c) and silicon ($\xi$=0.92±0.06 and $\Phi_{thr}^{(exp)}(N = 1) = 0.16 \pm 0.05$ J/cm$^2$) (Fig.1d). The error bars in both Figs. 1c and d, which are derived from the fitting on Figs. 1a and b respectively, are larger at lower number of pulses (*N*=22, 30 and 50). As shown in Figs. 1a and b, the measured values (dots) with the higher distance from the fitted curves are those for low number of pulses and low laser fluences. This is probably because at low number of pulses and fluences close to the ablation threshold, the edges of the ablated lines are not well defined at the edges due to the lower dose at the edges of the Gaussian profile laser beam, affecting the measurement of the ablated width. This results in higher deviation on the plotted values for low N in Figs. 1c and d. The above expression infers that for both silicon and nickel incubation effects are present, which yields an initially high value of the damage threshold at *N*=1. It is evident that the expected initial drop of the damage threshold at small number of pulses is expected to have significant implications in laser-based patterning. By contrast, at higher energy doses, the damage threshold reaches a plateau (Figs. 1c, d). It is noted that despite the lack of experimental results for smaller *N* values the evolution of the damage threshold curve is assumed to follow physical processes that result from the occurrence of incubation effects at small energy doses (i. e. small *N*). Nevertheless, a more comprehensive investigation is required to elucidate the likelihood of the absence of such effects at extremely small pulse durations (<50 fs). This assumption is generated from results in previous reports which show that for very short pulses (~5 fs), the ablation threshold is independent of the pulse number [62].

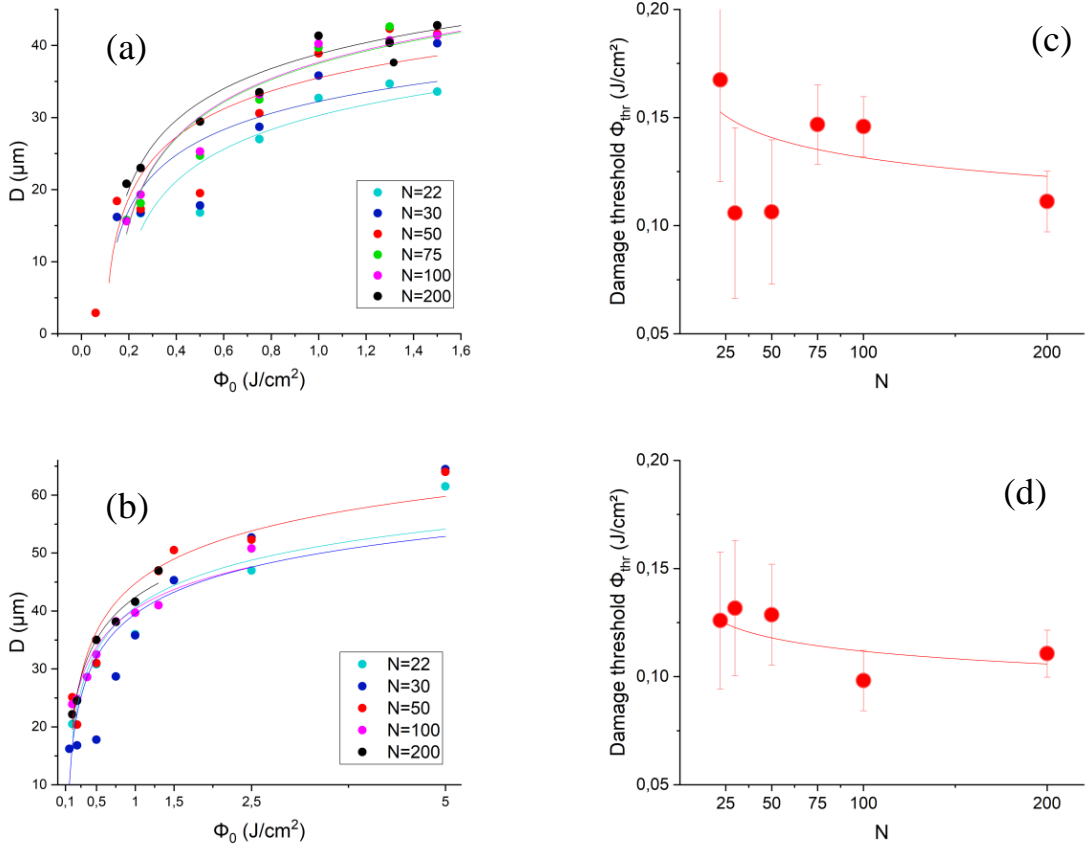

Figure 1: Measured diameters of ablated region for different *N* for nickel (a) and silicon (b). Experimentally calculated damage thresholds as a function of the number of pulses for nickel (c) and silicon (d) ($\tau_p$=45 fs). *Red* line in (c), (d) results from the fitting of experimental data (*red* dots).

To correlate the experimental results with the simulated values of the damage threshold, a thorough investigation of the ultrafast dynamics and calculation of the fluence inducing minimum surface modification are required. The minimum fluence for the generation of a permanent change on the solid's surface is taken as a criterion for the damage threshold $\Phi_{thr}$ and, in principle, there is an ambiguity on the determination of $\Phi_{thr}$. More specifically, a morphological change can be induced (i) either through mass redistribution (with conservation of the total mass) originated from a phase transition, melting and resolidification of a portion of the material [10] or (ii) through mass removal (with loss of material) due to ablation [63]. Thermal criteria are used in both cases to determine the onset of damage; thus, $\Phi_{thr}$ is defined as the minimum fluence required to melt the material (i.e. to raise the lattice temperature above the material's melting point, which equals 1685 K and 1728 K, for silicon [10] and nickel [16], respectively), i.e. Temperature above Melting Point, TMP; or to force the material to undergo a phase Transition to a Superheated Liquid (TSL) (i.e. the lattice temperature reaches $0.90 \times T_{cr}$, where $T_{cr}$ stands for the critical point of the material, i.e. 9460 K and 5159 K, for nickel [64] and silicon [10], respectively) [63].

**a. Nickel**

Without a loss of generality, the TMP criterion is used in this study to evaluate the damage threshold in nickel. To reveal the role of the laser parameters in the damage threshold, special emphasis is placed on the impact of the pulse duration. To this end, an analysis has been performed to investigate the ultrafast dynamics and the minimum laser fluence at various pulse durations (Fig.2a). It appears that an initial increase of the damage threshold is followed by a drop at ~1 ps before the threshold value starts to rise again at ~14 ps. To interpret the physical events at which these changes occur, it is important to evaluate the interplay between two processes through which hot electrons on the surface lose energy, namely, diffusion inside the volume by means of diffusion and electron-phonon scattering. Therefore, the temperature variation is firstly calculated at a relatively small fluence, 0.15 J/cm² (Fig.2b), which demonstrates that at pulse durations shorter than the approximately estimated time required for electron-phonon relaxation $\tau_{eq}$ ($\tau_{eq}$ ~ 8 ps), electron diffusion represents the dominant factor that leads to electron energy loss. More specifically, highly energetic electrons move deeper inside the volume, which indicates that at increasing electron diffusion, the energy transfer to the lattice system decreases. As a result, a maximum value of $T_L$ occurs near $\tau_{eq}$, where electron-phonon scattering starts to dominate. A similar peak value for $T_L$ at $\tau_p \sim \tau_{eq}$ was presented in a previous report [65]; furthermore, through electron-phonon coupling, the produced $T_L$ is expected to drop at decreasing $\tau_p$ (for $\tau_p<\tau_{eq}$) given the increase of electron thermal conductivity at moderately high electron temperatures. Interestingly, at even smaller values of $\tau_p$ ($\tau_p<<\tau_{eq}$) the maximum lattice temperature starts to increase (Fig.2b). This behaviour can be attributed to the large electron temperatures that are attained in those conditions (Fig.2c) that gradually lead to a drop of electron thermal conductivity (Fig.2d) at high temperatures. Thus, energy transfer from the electron to the lattice subsystems is hindered, which results in an increase of $T_L$ (Fig.2b). The above discussion for temperature evolution justifies the trend for the damage threshold's dependence on $\tau_p$ (i.e. as it should follow an inverse trend to the $T_L$ behavior). Notably, for higher fluences, the pulse duration value at which the extrema of the damage threshold (or equivalently the lattice temperatures) occur is shifted to larger values compared to the ones calculated for 0.15 J/cm². This is due to the fact that at higher fluences the larger electron temperatures that are produced lead to a smaller electron-phonon coupling efficiency for nickel [66] and therefore, the electron-phonon relaxation is delayed.

The above analysis for the conditions of the experiments in this work ($\tau_p$=45 fs) predicts a theoretical value $\Phi_{thr}^{(simul)}(N=1) = 0.315$ J/cm² for a single pulse, which is within the experimental error of the measured value (i.e. $0.2 \pm 0.1$ J/cm²).

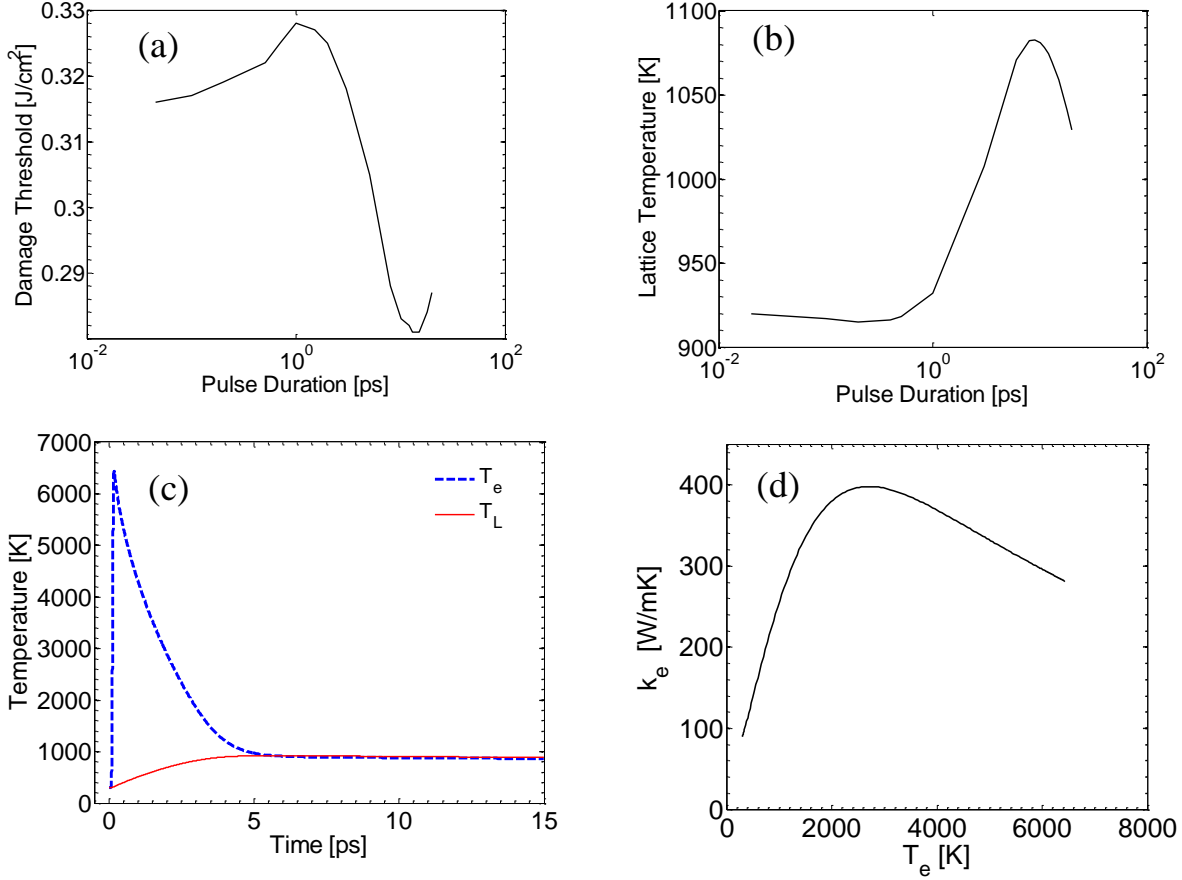

Figure 2: Simulations for nickel: (a) Damage threshold *vs* pulse duration (single pulse), (b) Lattice temperature dependence on pulse duration for 0.15 J/cm$^2$, (c) Electron and lattice temperature evolution (for 0.15 J/cm$^2$, $\tau_p$=45 fs), (d) Electron thermal conductivity as a function of electron temperature.

**b. Silicon**

The dependence of the damage threshold for silicon following a single shot irradiation of the solid was evaluated in a previous report at various laser wavelengths in the mid-IR region [46]. According to that analysis, the theoretical value of the damage threshold scales as $\Phi_{thr}^{(simul)}(N=1) \sim \tau_p^{0.552}$ at $\lambda_L$=3200 nm. The theoretical model for the calculation of the damage threshold in various laser conditions at mid-IR agrees well with experimental results [35, 46]. However, our results demonstrate a discrepancy between the measured value and the theoretical prediction $\Phi_{thr}^{(simul-melt)}(N=1) \sim 0.070$ J/cm$^2$. It is noted though that compared to the rTTM theoretical model used for metals, the model applied to silicon does not include the contribution of the out-of-equilibrium electron population. On the other hand, previous papers [49, 50], reported that using rTTM affects the calculated maximum lattice temperature (i.e. the nonthermal electrons influence the resulting lattice temperature). This suggests that a possible interpretation could be attributed to the role of nonthermal electrons, but the investigation of this interpretation is beyond the scope of the current study. Furthermore, while the TMP value constitutes a reasonable criterion to estimate the damage threshold, the fulfilment of the requirement to produce even a very thin layer of molten volume to ensure a phase transition and damage is difficult to verify experimentally; in principle, a very shallow molten volume is difficult to be measured (i. e. TMP is related to the onset of a phase transition *on* the surface). Thus, to produce a sufficiently large and measurable volume of the size of a few nanometers, a larger fluence has been used, which leads to mass removal. The theoretical prediction for the TSL-based criterion yields a value $\Phi_{thr}^{(simul-ablation)}(N = 1) \sim 0.18$ J/cm². The above discussion indicates that the damage threshold is reasonably assumed to be between 0.070 and 0.18 J/cm², which is within the experimental error of the measured value (i.e. $0.16 \pm 0.05$ J/cm²).

## II. Surface Patterning

To evaluate the influence of the laser conditions on LIPSS formation, we have performed a systematic analysis to reveal the impact of two parameters, fluence and energy dose, which are usually employed to explore laser patterning. Figs. 3 and 4 present a detailed parametric study of the various types of LIPSS formed on silicon and nickel, respectively, following irradiation with linearly polarized mid-IR femtosecond laser pulses.

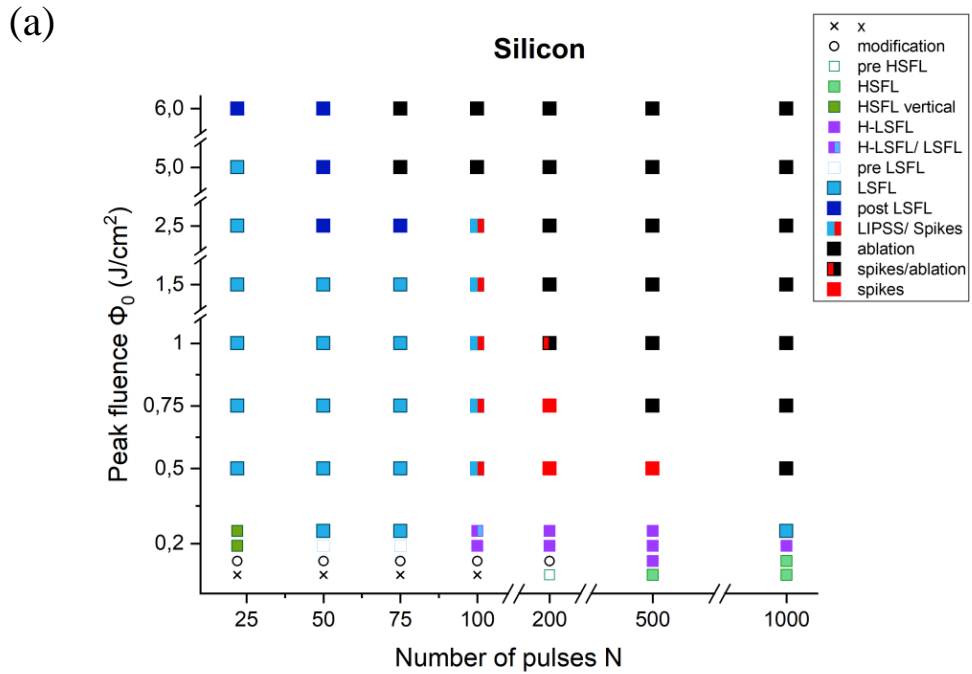

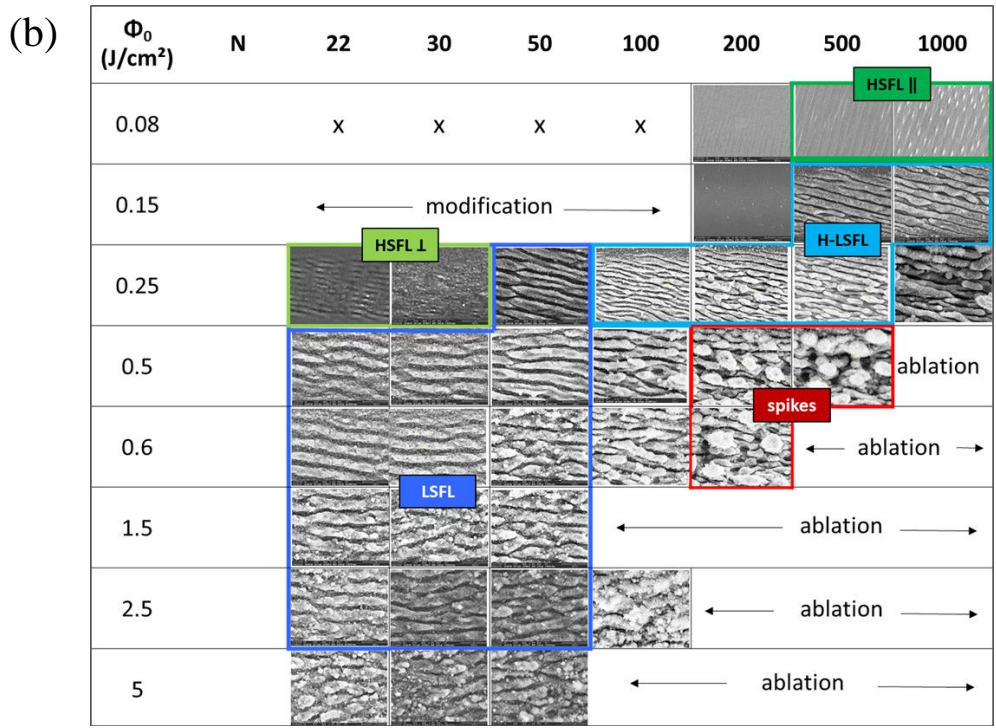

Figure 3: Map of morphologies obtained on silicon surfaces upon irradiation with a series of laser fluences and number of pulses (a), (b). The colour symbols indicate the different types of surface topographies. The corresponding SEM micrographs of the microstructures observed are presented in (b). In all such images, the direction of the laser polarization is vertical.

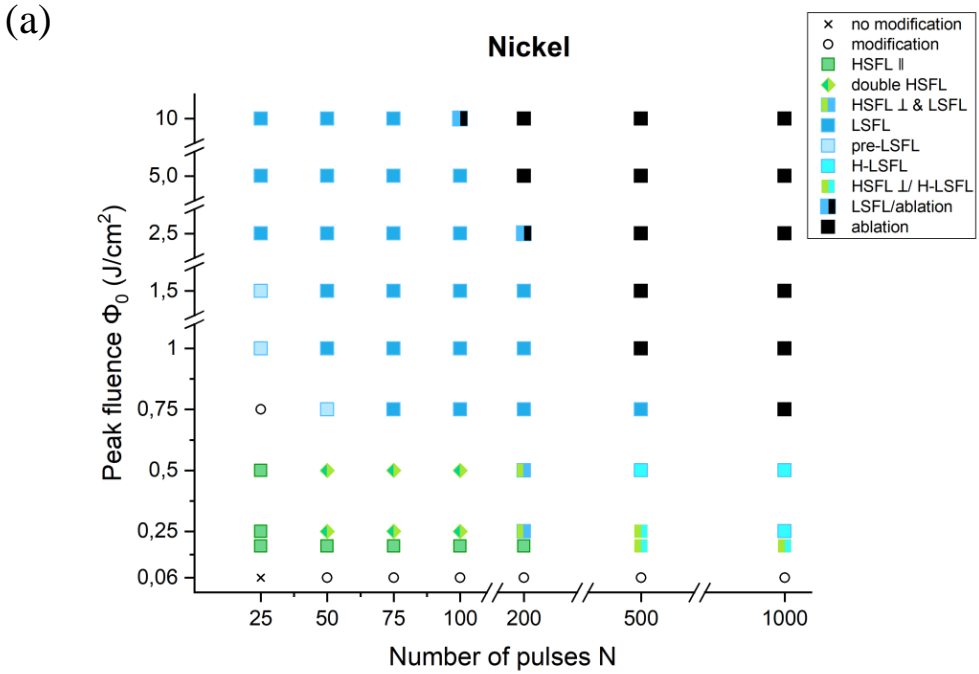

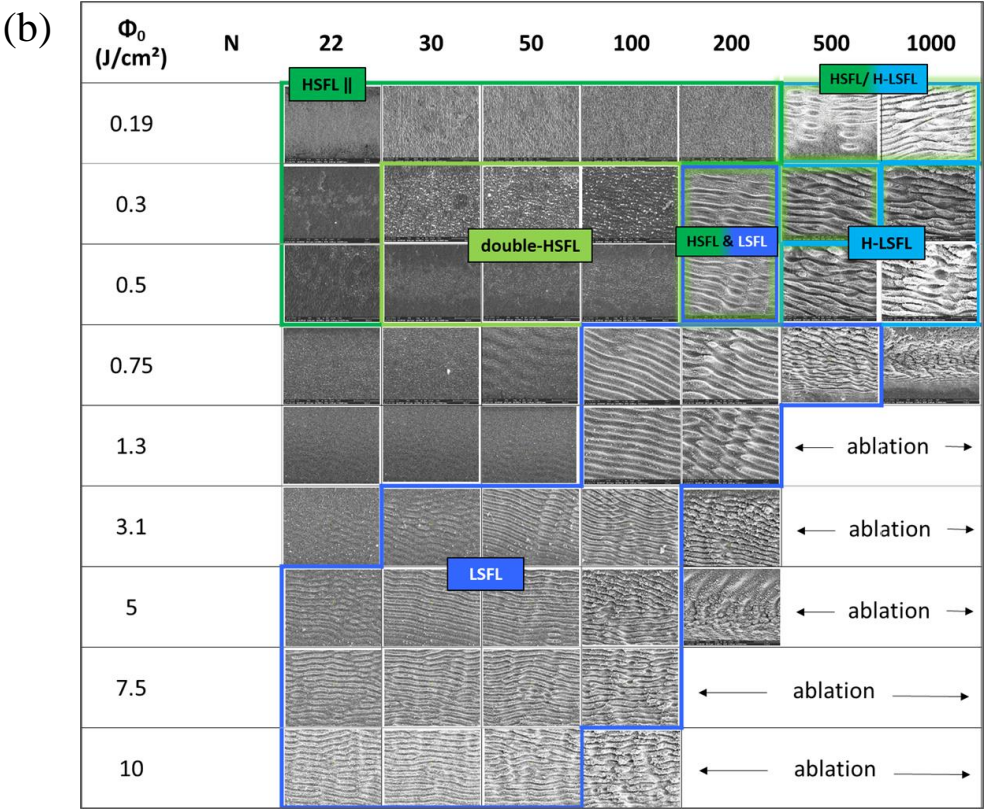

Figure 4: Map of morphologies obtained on nickel surfaces upon irradiation with a series of laser fluences and number of pulses. The colour symbols indicate the different types of surface topographies (a), (b). The corresponding SEM micrographs of the microstructures observed are presented in (b). In all such images the direction of the laser polarization is vertical.

In particular, the upper graph of each figure (i. e. Figs. 3a, 4a) shows a map of the various types of microstructures observed under different irradiation conditions. The corresponding morphological characteristics of each microstructure type are illustrated in the *lower* images (i. e. Figs. 3b, 4b) in the form of SEM micrographs. A rough comparison between the semiconductor (silicon) and the metal (nickel) indicates a bigger variety of microstructures for Si. On the other hand, LSFL on Ni are formed over a larger window of laser fluences and number of pulses. A more detailed investigation of the topographies of each material is discussed below.

**a. Silicon**

The results in Fig. 3 indicate a variety of periodic surface structures observed on Si upon irradiation with 3.2 μm fs laser pulses. In particular, LIPSS with very low periodicities, the so-called HSFL, appear with different orientation, depending on the number of pulses, while their periods range from $\lambda_L/10$ to $\lambda_L/5$. HSFL with orientation parallel to the laser polarization are observed for low energies and high number of pulses ($N \geq 500$ at 0.08 J/cm²), while HSFL perpendicular to the polarization are formed at lower number of pulses ($N = 22$–30 at 0.25 J/cm²). In general, HSFL perpendicular to the polarization are formed upon irradiation with a much lower laser dose (5.5–7.5 J/cm²), resulting in more shallow structures, in comparison to the parallel HSFL (40–80 J/cm²). On the other hand, LSFL are observed in a broad window of fluences and pulses. Particularly, LSFL with periodicities close to the laser wavelength ($\Lambda \approx 2.5$ μm) appear at low number of pulses ($N \leq 50$) and a wide range of laser fluences (0.25–2.5 J/cm²). Interestingly, at similar doses but higher number of pulses ($N \geq 100$), LSFL with much lower periodicities, close to $\lambda_L/2$ ($\Lambda \approx 1.5$ μm) are formed, which are marked in the plot (Fig. 3) as high-LSFL (H-LSFL). Finally, microspikes are formed at low fluences, but high number of pulses. The periodicity dependence on the laser polarization direction is shown in more detail in Fig. 5. To understand the formation of LIPSS on silicon, following irradiation with mid-IR pulses, the excitation level and surface corrugation features are usually considered to determine the orientation as well as the periodicity of the induced LIPSS upon repetitive irradiation (i.e. at increasing energy dose). With respect to the underlying physical mechanisms that account for the formation of LSFL structures, a comparison of electromagnetic effects and electron density variations between IR and mid-IR irradiation and at different fluences were presented in previous reports for semiconductors [46] and dielectrics [45]. To correlate $N_c$ with the orientation and periodicity of LSFL structures, we use Sipe's theory, which is based on the interference of the incident laser beam with some form of surface-scattered electromagnetic waves [67, 68]. More specifically, the inhomogeneous energy deposition into the material is computed through the calculation of the product $\eta(\vec{K},\vec{\kappa_\iota})\times|b(\vec{K})|$, in which $\eta$ stands for the efficacy with which surface roughness at the wave vector $\vec{K}$ (i.e., normalized wave vector $|\vec{K}|=\lambda_L/\Lambda$, where $\Lambda$ stands for the predicted structural periodicity), induces inhomogeneous radiation absorption, $\vec{\kappa_\iota}$ is the component of the wave vector of the incident laser beam on the material's surface plane and *b* represents a measure of the amplitude of the surface roughness at $\vec{K}$. It is noted that compared to the standard values for the 'shape', *s*, and the 'filling', *f*, parameters (i.e. 0.4 and 0.1, respectively) [67], roughly shaped islands and particular values for *s*, *f* (i.e. 0.4 and 0.7, respectively) are used to calculate the efficacy factors and $\Lambda$ at different values of $N_c$ (i.e. $N_c=0.55\times10^{21}$ cm$^{-3}$ (Fig. 6a) and $N_c=4\times10^{21}$ cm$^{-3}$ (Fig. 6b)). According to Sipe's theory, sharp points of $\eta$ appear along the $K_x$ $(=\lambda_L/\Lambda_x)$ for $N_c=0.55\times10^{21}$ cm$^{-3}$, which indicates that periodic structures oriented perpendicularly to the laser polarization are produced with $\Lambda=\Lambda_x$ close to the periodicity of the laser wavelength. By contrast, $\eta$ is very small for $N_c=4\times10^{21}$ cm$^{-3}$, which suggests that LSFL should not be observed at very high excitation levels. Based on Sipe's theory, the periodicity of the induced surface scattered waves varies as a function of $N_c$, which resembles the evolution trend characteristic of those waves that are assumed to result from the excitation of SP (Fig. 6c). A similar conclusion has been deduced to explain the periodicity of surface waves upon the irradiation of solids with IR [69] or dielectrics with mid-IR pulses [45]. A projection of the impact of the (peak) fluence value on $\Lambda$ is presented in Fig. 6d; this explains the increase of the period through the production of higher excitation levels as the fluence becomes larger. Although these results correspond to irradiation with *N*=2, the monotonically increasing $\Lambda$ shown in Fig.6d is valid at higher energy doses, *N*=22 (Fig. 7a).

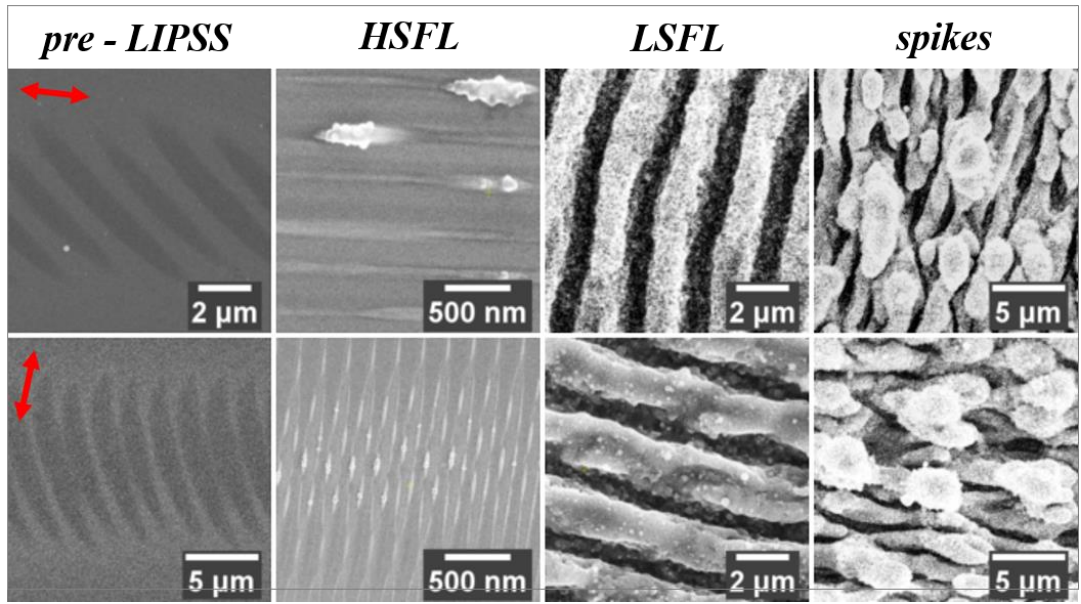

Figure 5: SEM micrographs of periodic surface structures on silicon upon mid-IR fs laser pulse irradiation. The double arrow on the left indicates the laser polarization direction.

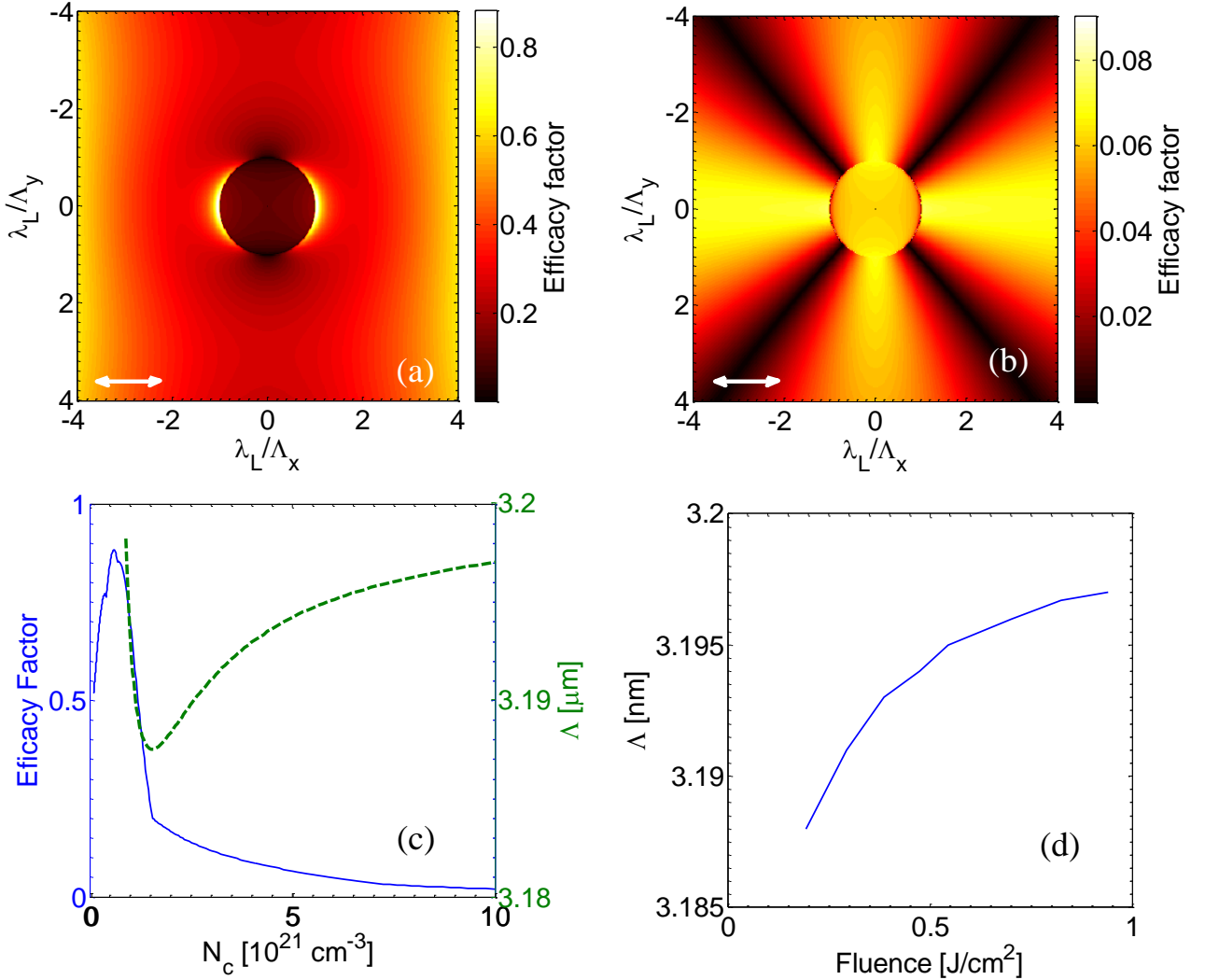

Figure 6: Efficacy factor map for (a) $N_c=0.55\times10^{21}$ cm$^{-3}$, (b) $N_c=4\times10^{21}$ cm$^{-3}$ (the *white* double arrow indicates the orientation of the laser polarization), (c) Efficacy factor values and the corresponding $\Lambda$ as a function of $N_c$, (d) LSFL periodicity as a function of fluence.

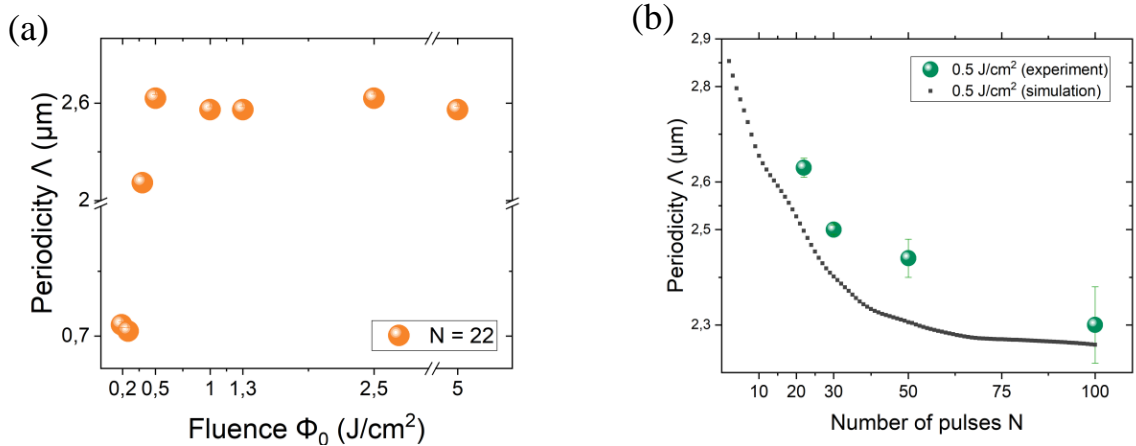

Figure 7: The evolution of the ripples' periodicity on a silicon surface, (a) as a function of laser fluence at *N*=22 and (b) as a function of the number of laser pulses at $\Phi_0 = 0.5$J/cm². The *dotted* line in (b) represents the theoretical predictions.

Although the electromagnetic fingerprint dictates the frequency of the induced rippled pattern, the laser energy and thermal effects play a very significant role in the final morphology attained. Previous works reported that the interference of the scattered wave with the incident beam produces an inhomogeneous intensity profile that modulates the thermal response of the electron/lattice subsystems; a resulting periodic modulation of the lattice temperature of the solid and the produced thermocapillary effects lead eventually to rippled topographies (see Refs [8, 10, 16] and references therein). Theoretical results for the evolution of the LSFL periodicity as a function of *N* are shown in Fig. 7b, which demonstrates good agreement with the experimental observations. The evaluation of the ripple periodicity was based on an approximating methodology that allows the calculation of the SP wavelength as a function of the depth of the topography and the features of the corrugated profile that is produced when the energy dose increases [70]. A more precise approach should reveal the impact of electromagnetic effects through simulations based on the investigation of light scattering off a rough or patterned surface [8, 71]. Simulations for the formation of subwavelength structures perpendicular to the laser polarization are illustrated for *N*=2 and 0.5 J/cm² (peak fluence) in Fig. 8, which shows the induced topography.

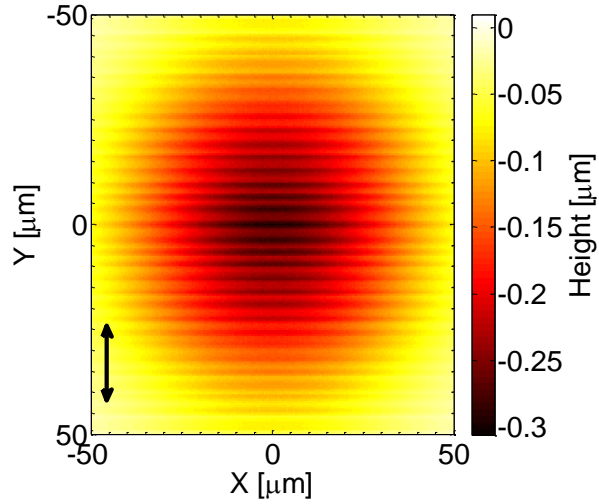

Figure 8: Surface pattern for *N*=2 (0.5 J/cm$^2$, the *black* double arrow indicates the laser polarization orientation).

The experimental results shown at the beginning of the section indicate that LIPSS similar to those fabricated at lower spectral regions [1, 72] are formed in various conditions at mid-IR, however, compared to the observations for IR pulses, it appears that no microgrooves are produced. More specifically, at

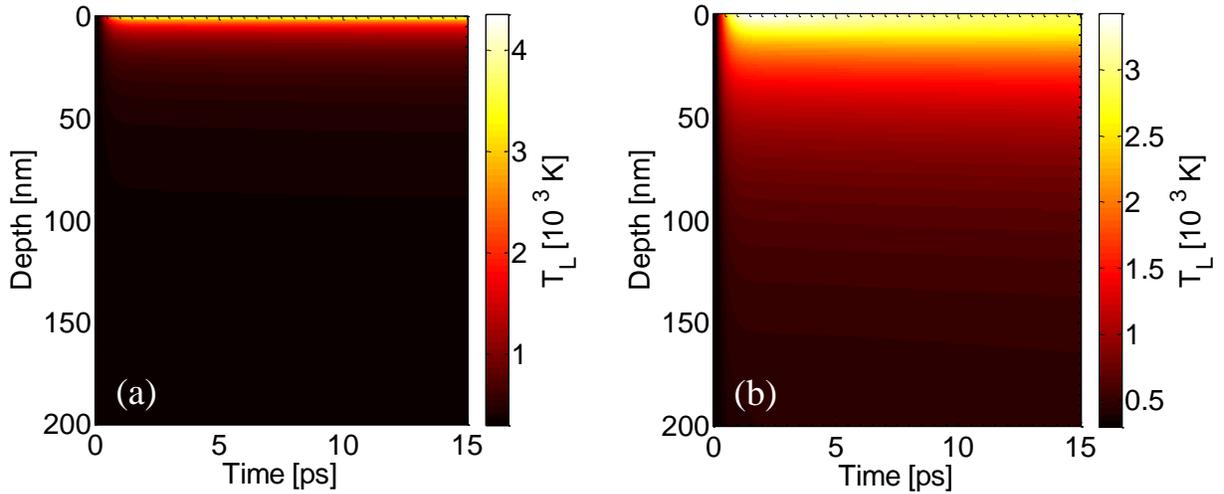

Figure 9: Spatiotemporal evolution of the lattice temperature following irradiation with a single shot for (a) $\lambda_L$=3200 nm, (b) $\lambda_L$=800 nm ($\Phi$=0.15 J/cm$^2$, $\tau_p$=45 fs).

lower – IR – laser wavelengths, a transition from LSFL structures to suprawavelength grooves was observed and predicted theoretically, prior to the formation of spikes [13]; by contrast, at $\lambda_L$=3200 nm, the experimental results do not justify the production of grooves. Given that a deep corrugated profile is required to facilitate a preferential direction of the fluid transport that will lead to the formation of grooves, an analysis of the remarkably different impact on the thermal effects of mid-IR pulses appears to be sufficient to partly explain the absence of such structures. A thorough investigation of the size of the produced molten region for mid-IR (Fig.9a) and IR pulses (Fig.9b), simulations demonstrate a deeper molten volume for IR pulses. More specifically, for a peak fluence of 0.15 J/cm$^2$, the spatiotemporal evolution of the temperature field shows that the molten region extends to a depth equal to ~6 nm (for $\lambda_L$=3200 nm) compared to ~35 nm (for $\lambda_L$=800 nm). Although these 1D simulations reflect the thermal fingerprint of the laser pulses for a single shot, similar conclusions can be deduced at higher energy doses and for 3D profiles as an enlargement of the depth difference is expected at the two spectral regions. It is evident that a different combination of $N$ and fluence values could lead to the formation of grooves if appropriate molten volume conditions are satisfied, however, the values used in the experiments in this work did not evidence the formation of such suprawavelength structures. On the other hand, a theoretical analysis of the convection rolls development and movement that operate as the precursors for groove formation show that parameters such as the molten volume size and the Marangoni number [13] play a very important role in the production of thermocapillary waves that will eventually lead to stable periodic structures. As some preliminary theoretical results do not suggest that groove formation is inhibited, a more systematic experimental investigation is required to reveal whether there is a range of experimental values at which suprawavelength structures can be produced. It is noted that similar approaches to Ref [13] for groove formation which are attributed to capillary waves have been presented in other reports [73, 74]; the period of these suprawavelength structures are calculated through the dispersion relation for capillary waves in a shallow liquid layer. It must, also, be emphasized that the formation of spikes has been linked to the prior development of the suprawavelength structures for IR pulses [13]. Thus, if grooves are not formed, a fundamental question is whether another physical mechanism should be introduced to account for the development of the observed spikes for irradiation with pulses in the mid-IR spectral region. This is a very important aspect that should be explored in more detail in order to consistently describe the transition between all types of the produced structures.

Finally, the experimental results (see Figure 10) indicate that at high number of pulses and at very low fluence values, shallow periodic structures are formed that have the characteristics of HSFL structures and they are oriented parallel to the laser beam polarization. Such double-periodic structures are different from LIPSS, and therefore – although the mechanism of their formation is interesting and is being investigated – they are beyond the scope of this work.

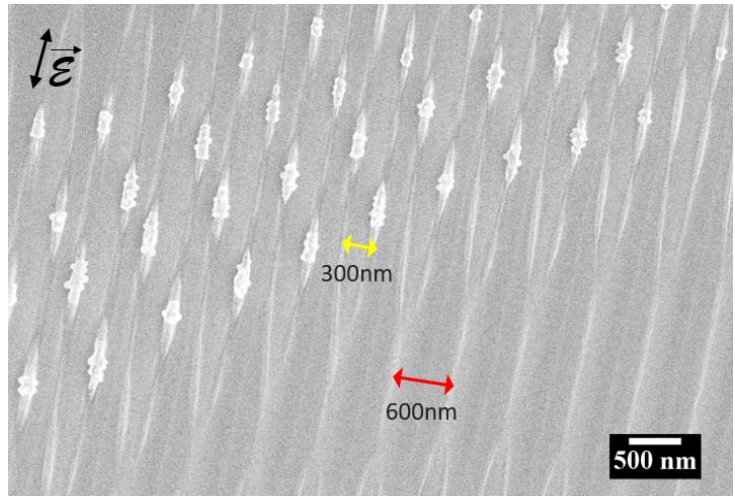

Figure 10: HSFL LIPSS on Si upon irradiation with $N$=1000 laser pulses at 0.1 J/cm² laser fluence. Two different periodicities equal to $\lambda_L$/10 and $\lambda_L$/5 appear within the same area as indicated by the arrows on the image. HSFL structures are oriented parallel to the polarization.

**b. Nickel**

The different types of periodic structures observed on nickel upon irradiation with 3.2 μm infrared fs laser pulses are reported in Fig. 4. HSFL appear at low fluences ($\Phi_0 = 0.19$-$0.5$ J/cm²) and $N \leq 200$. It is worth mentioning that the range of irradiation conditions at which LSFL are observed is much wider for nickel than for silicon. And in the case of silicon, HSFL (or their equivalent, i.e. LSFL with a much lower periodicity close to $\lambda_L/2$) are also formed at high number of pulses ($N \geq 500$) for low laser fluences.

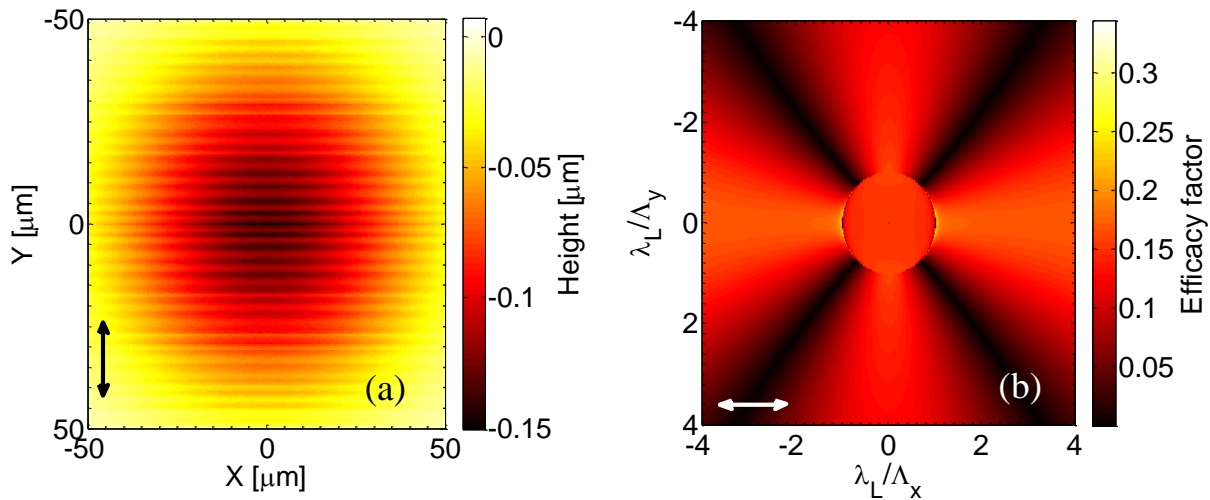

Figure 11: (a) Surface pattern for $N=2$ (0.5 J/cm²), (b) efficacy factor map (double arrow indicates the laser polarization orientation).

A computational approach similar to that used for silicon can be performed to predict LSFL formation on nickel surfaces. Fig.11a illustrates the LSFL topography on nickel. It shows that SP excitation leads to the formation of periodic structures of frequencies determined by the SP wavelength. Alternatively, by applying Sipe's theory assuming that the dielectric parameter of nickel is equal to $\varepsilon=-125+88.3i$ at 3200 nm [75], simulations show the formation of LSFL structures perpendicular to the laser polarization (Fig.11b) with a periodicity similar to that predicted from the SP theory. Other morphological features of the LSFL (i.e. height) are predicted through the employment of the multiscale physical model presented in the previous sections (Eqs.1 and 5).

With respect to LSFL's periodicity dependence on fluence, Fig. 12a demonstrates monotonically increasing periodicity of LSFL as the laser fluence becomes higher. At the same time, LSFL spacing decreases as the number of pulses increases as shown in Fig. 12b. Simulation and experimental results at 0.5 J/cm² demonstrate a good agreement in the trend of the periodicity decrease at higher $N$, while there is an obvious discrepancy in the periodicity values. In a previous report, it was shown that in order to correlate experimental results with a consistent theoretical model, appropriate changes are required to be made in the dielectric parameter by incorporating the impact of the fluence on the electron collision frequency and the plasma resonance as well as the role of the interband transitions and hot electron localization [76]. Nevertheless, as nonthermal electrons are expected to form during the pulse, and the instantaneous thermalization of the electronic distribution is not possible, a more complex computational framework is required to provide a more precise evaluation of the transient optical properties. Such theoretical approaches could involve the employment of density functional theories [77, 78]. Therefore, a revised simulation tool accounting for the ultrafast dynamics of out-of-equilibrium electrons (Eq.1) should include a component that describes the variation of the optical properties of the irradiated material and that of the absorbed energy in detail. Apart from the capacity to estimate precisely the correlation of the periodic structure features with the laser parameters, the theoretical approach could reveal the significant role of the out-of-equilibrium electrons in the early stages of irradiation.

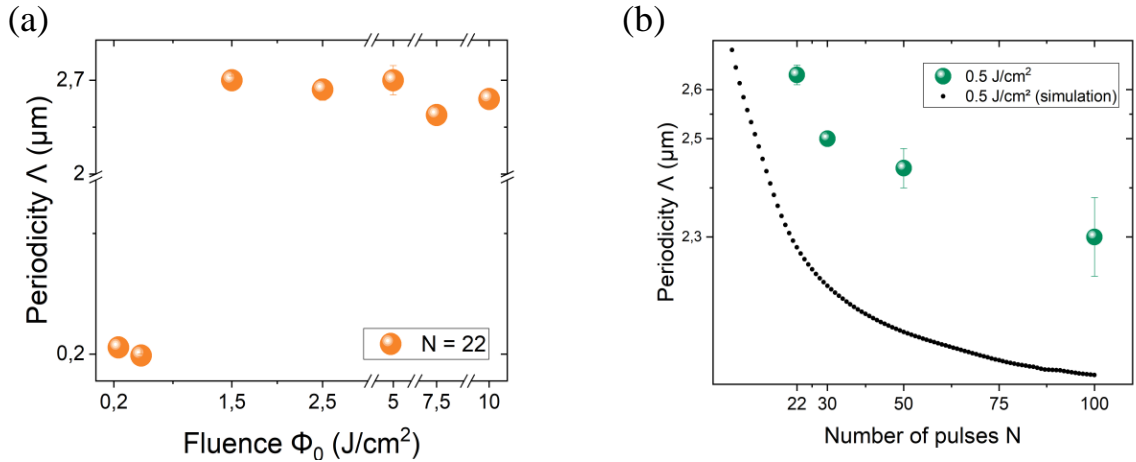

Figure 12: The evolution of the ripples periodicity for nickel (a) as a function of the laser fluence at $N$=22 and (b) as a function of the number of laser pulses at $\Phi_0 = 0.5$ J/cm². The *dotted* line in (b) represents the theoretical predictions.

Finally, experimental results (Figure 13) indicate that at high number of pulses and at low fluence values (below the ablation threshold) one sees the formation of shallow periodic structures that have characteristics typical of HSFL structures and are oriented perpendicular to the polarization of the laser beam. Interestingly, at such laser doses, ripples with three different periodicities coexist on the surface as shown in Fig. 13b where the periodicities measured through 2D-FFT are: $2.77 \pm 0.08$ μm, $1.31 \pm 0.01$ μm and $0.221 \pm 0.003$ μm. Fig 13c represents a higher magnification of Figure 13b, where the HSFL are visible. All of them are oriented *perpendicular* to the laser polarization and they could be classified as LSFL, High-LSFL (H-LSFL) and HSFL, respectively, according to their size. The structures that are coined as H-LSFL are those that have spatial periodicities between the periodicities of LSFL and HSFL. Those periodicities are shortly below $\lambda_L/2$. In the literature, ripples of such a size are also classified as HSFL [7]. In any case, three different types of ripples appear on the surface (Figure 13b, c): (i) LSFL with a size of 2.77 µm (ii) large HSFL at 1.3 µm and (iii) HSFL at 0.22 µm. The exact mechanism for these HSFL formations is not yet clear. According to the results in previous reports, HSFL formation on metallic surfaces, perpendicular to the polarization can be partly attributed to twinning effects during the resolidification of a shallow laser-induced melt layer [79]. This could explain our results, as the structures appear at very low fluence values. Nevertheless, further investigation is required in this field.

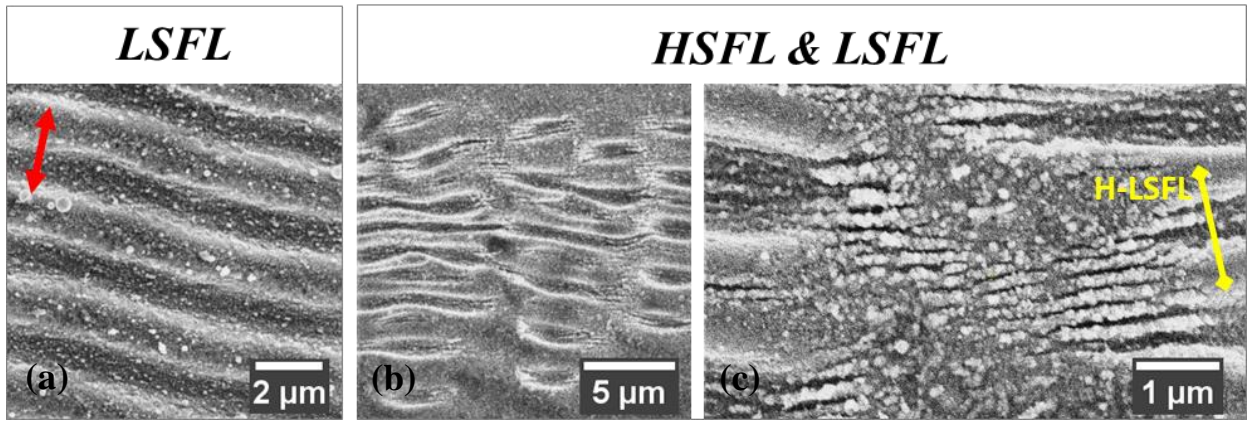

Figure 13: Different types of ripples on nickel upon irradiation with mid-IR fs pulses: (a) Typical LSFL appear for 1.3 J/cm² and $N$=100 with a periodicity close to the laser wavelength ($\Lambda \approx 2.8$ μm), while (b) at higher number of pulses and fluence values almost below the ablation threshold (0.19 J/cm², $N$=500) more complex ripples appear, where LIPSS with three different periodicities co-exist. (c) A magnification of (b) shows more clearly the H-LSFL structures with 1.3 μm periodicity and the HSFL with 0.2 μm spatial periods.

## E. Conclusions

Surface patterning with ultrashort mid-IR laser pulses was systematically explored for silicon and nickel, and the conditions for the formation of a variety of LIPSS were investigated. Interestingly, for silicon, while most of the LIPSS developed at lower wavelengths (HSFL, LSFL, spikes) were also formed by mid-IR pulses, experimental results indicated the absence of groove formation, which was attributed to a predicted shallow molten volume that is produced at this spectral region. By contrast, on nickel surfaces, in addition to the formation of LSFL and HSFL structures, another type of structure with periodicities between those of LSFL and HSFL (H-LSFL) is produced. On the other hand, the damage threshold evaluation for nickel showed that at small pulse durations an initial increase of the damage threshold is followed by a drop at ~1 ps before the threshold value starts to rise again at ~14 ps, which is explained by the timescales at which electron diffusion or electron-phonon scattering become important. These results are presented through a combined experimental parametric study and a theoretical interpretation and they are aimed to provide the basis for a novel technique for the surface engineering of solids with strong mid-IR fields.

**CRediT authorship contribution statement**

**Stella Maragkaki:** Conceptualisation, methodology, validation, formal analysis, investigation, writing-original draft, **George D Tsibidis:** Conceptualisation, methodology, theoretical model and software, investigation, formal analysis, writing-original draft, funding acquisition, **L. Haizer:** data acquisition, **Z. Pápa:** data acquisition, **R. Flender:** data acquisition, **B Kiss:** data acquisition, **Z. Márton:** data acquisition, writing-review and editing, **Emmanuel Stratakis:** Conceptualisation, supervision, writing-review and editing, funding acquisition

**Declaration of competing Interest**

The authors declare that they have no known competing financial interests or personal relationships that they could have appeared to influence the work reported in this paper.

**Acknowledgements**
The authors would like to acknowledge the support provided by the European Union's Horizon 2020 research and innovation program through the *BioCombs4Nanofibres* project (Grant Agreement No. 862016). GDT acknowledges support from COST Action *TUMIEE* (supported by COST-European Cooperation in Science and Technology). The ELI ALPS project (GINOP-2.3.6-15-2015-00001) is supported by the European Union and co-financed from the European Regional Development Fund. The authors acknowledge the contribution of Krishna Murari from ELI-ALPS during the surface patterning work and are indebted to Judit Zelena for thorough proof reading.